\documentclass[journal]{IEEEtran}
\usepackage{times}

\usepackage{soul}
\usepackage{float}
\usepackage{algpseudocode}
\usepackage{algorithm}
\usepackage{url}
\usepackage[hidelinks]{hyperref}
\usepackage[utf8]{inputenc}
\usepackage{amsmath}
\usepackage{booktabs}
\usepackage{amsfonts}
\usepackage{bm}
\usepackage{amsmath} 
\usepackage{enumerate}
\usepackage{diagbox}             
\usepackage{multirow}  
\usepackage{makecell}
\usepackage{bbm}
\usepackage{threeparttable}

\usepackage{cite}

\ifCLASSINFOpdf
  \usepackage[pdftex]{graphicx}
\else
   \usepackage[dvips]{graphicx}
\fi
\ifCLASSOPTIONcompsoc
 \usepackage[caption=false,font=normalsize,labelfont=sf,textfont=sf]{subfig}
\else
 \usepackage[caption=false,font=footnotesize]{subfig}
\fi

\ifCLASSOPTIONcaptionsoff
 \usepackage[nomarkers]{endfloat}
\let\MYoriglatexcaption\caption
\renewcommand{\caption}[2][\relax]{\MYoriglatexcaption[#2]{#2}}
\fi
\begin{document}
%
\title{Sound Event Detection Transformer:\\ An Event-based End-to-End Model for Sound Event Detection}
%
%
%
\author{
Zhirong Ye$^{1,2}$, Xiangdong Wang$^{1,\dagger}$, Hong Liu$^1$, Yueliang Qian$^1$, Rui Tao$^3$, Long Yan$^3$, Kazushige Ouchi$^3$

\thanks{$^1$Beijing Key Laboratory of Mobile Computing and Pervasive Device, Institute of Computing Technology, Chinese Academy of Sciences, Beijing, China}
\thanks{$^2$University of Chinese Academy of Sciences, Beijing, China}
\thanks{$^3$Toshiba China R\&D Center, Beijing, China}}

%
%

\markboth{Journal of \LaTeX\ Class Files,~Vol.~14, No.~8, August~2015}%
{Shell \MakeLowercase{\textit{et al.}}: Bare Demo of IEEEtran.cls for IEEE Journals}
%



\maketitle

\begin{abstract}
Sound event detection (SED) has gained increasing attention with its wide application in surveillance, video indexing, etc. Existing models in SED mainly generate frame-level prediction, converting it into a sequence multi-label classification problem. A critical issue with the frame-based model is that it pursues the best frame-level prediction rather than the best event-level prediction. Besides, it needs post-processing and cannot be trained in an end-to-end way. This paper firstly presents the one-dimensional Detection Transformer (1D-DETR), inspired by Detection Transformer for image object detection. Furthermore, given the characteristics of SED, the audio query branch and a one-to-many matching strategy for fine-tuning the model are added to 1D-DETR to form Sound Event Detection Transformer (SEDT). To our knowledge, SEDT is the first event-based and end-to-end SED model.
Experiments are conducted on the URBAN-SED dataset and the DCASE2019 Task4 dataset, and both show that SEDT can achieve competitive performance. 
\end{abstract}


%
\IEEEpeerreviewmaketitle

\section{Introduction}
%
%
%
%
\IEEEPARstart{S}{ound} event detection (SED) is important for many applications, such as smart cities\cite{bello2018sonyc}, healthcare\cite{goetze2012acoustic}, surveillance\cite{crocco2016audio}, video indexing\cite{hershey2017cnn},  and so on. It consists of two subtasks, one is to recognize the event in an audio clip, and the other is to locate its corresponding start and end boundaries. Sound events in real life tend to overlap with each other considerably. Recognizing such overlapping sound events is referred to as polyphonic SED \cite{cakir2015polyphonic}. This is similar to the task of object detection in the field of computer vision whose goal is to locate and classify objects in an image where different objects may also be occluded and overlapped. Sound event detection and image object detection can be regarded as subtasks of object detection for one-dimensional and two-dimensional signals.

For SED, deep learning has shown initial success and became the mainstream. Existing work mainly employs convolutional neural networks (CNNs) \cite{zhang2015robust} or convolutional recurrent neural networks (CRNNs) \cite{turpault2019sound} to learn the representation for each frame, obtains the frame-level classification probability by a classifier, thresholds the probabilities to get frame-level decisions, and then poses post-processing such as median filtering to smooth the outputs. Recently, considering the superiority of the Transformer architecture \cite{vaswani2017attention}, Miyazaki et al. \cite{miyazaki2020convolution} propose to replace recurrent neural networks (RNNs) of the CRNN model with the encoder of Transformer or Conformer \cite{gulati2020conformer}. The above models transform sound event detection to a sequence-to-sequence multi-label classification problem.

However, a critical issue with the sequence-to-sequence classification scheme is the inconsistency between the goal of the model and the task, where the former pursues the best frame-level prediction while the latter aims at the best event-level prediction. Such inconsistency brings up two problems: First, the model is not optimal for the task, for its design and training are both for the best frame-level prediction. Specifically, for model architecture, to extract local information required by frame-level prediction, only a small time-axis compression scale is allowed, resulting in the failure of the model to capture the global information describing the event as a whole. For model training, the frame-level classification loss is adopted as the objective function. Hence the model obtained after training is compliant with the frame-level rather than the event-level target. In short, the frame-based model does not care about the correlation between frames, nor does it care about the event-level information, such as event duration, which counts for a great deal for event-level prediction. Second, for such frame-based models, post-processing is indispensable to aggregate frame-level prediction into event-level prediction. Some work \cite{miyazaki2020convolution, DCASE2019ICT} has greatly improved the performance of the model by setting class-adaptive window sizes and thresholds. However, this will introduce too many hyperparameters, resulting in difficulty in training and bringing the risk of over-fitting. Besides, these hyperparameters are specific to the dataset and need to be manually searched from scratch when applying to different datasets, which greatly reduces the generalization ability of the model. To resolve the inconsistency between the model target and the task, end-to-end learning has been proposed, which requires the model to output the final results directly to ensure that the model is optimized towards the task goal during training. This end-to-end philosophy has become a trend and achieves state-of-the-art (SOTA) results in many fields, such as speech recognition and machine translation, but not yet in SED. To construct an end-to-end SED model, an event-based prediction mechanism needs to be designed.

In image object detection, traditional models also only give intermediate outputs, and apply the non-max suppression (NMS) \cite{neubeck2006efficient} procedure to get the final results. However, the commonly adopted GreedyNMS cannot be easily parallelized and has become a performance bottleneck \cite{cai2019maxpoolnms}. To bypass the time-consuming post-processing, the first end-to-end object detection model, Detection Transformer (DETR) \cite{carion2020end}, has been proposed and achieves competitive performance. DETR implements an object-based prediction mechanism to directly predict the boundary and category label for each object without any post-processing.  Learning from the object-based prediction mechanism of DETR, we built to our knowledge the first end-to-end model for SED which outputs event-level predictions directly. The event-based prediction schema makes the model better adapted to the task.
For the event-based model, the time axis compression scale will be up to the event duration to take both local and global information into account to get better event representation. At the same time, it is capable of learning various event characteristics such as duration, without introducing hyperparameters during post-processing which are obtained by manual search on the development dataset.

In this paper, we propose to our knowledge the first event-based and end-to-end SED model. Firstly, the one-dimensional Detection Transformer (1D-DETR) is presented for SED, inspired by Detection Transformer for image object detection. And then, to overcome the two shortcomings we observed in the 1D-DETR model, namely, lacking category information and tending to judge events as the ``empty'' category due to the one-to-one matching principle adopted during training, we propose the audio query branch and the one-to-many matching strategy to obtain a Sound Event Detection Transformer (SEDT) which does not need any post-processing and realizes end-to-end sound event detection. Experiments on the URBAN-SED dataset and the DCASE2019 Task4 dataset both show that SEDT achieves competitive performances compared with SOTA models. The code will be available soon.

\section{Related work}
\begin{figure*}[!t]
\centering
\includegraphics[width=7.16in]{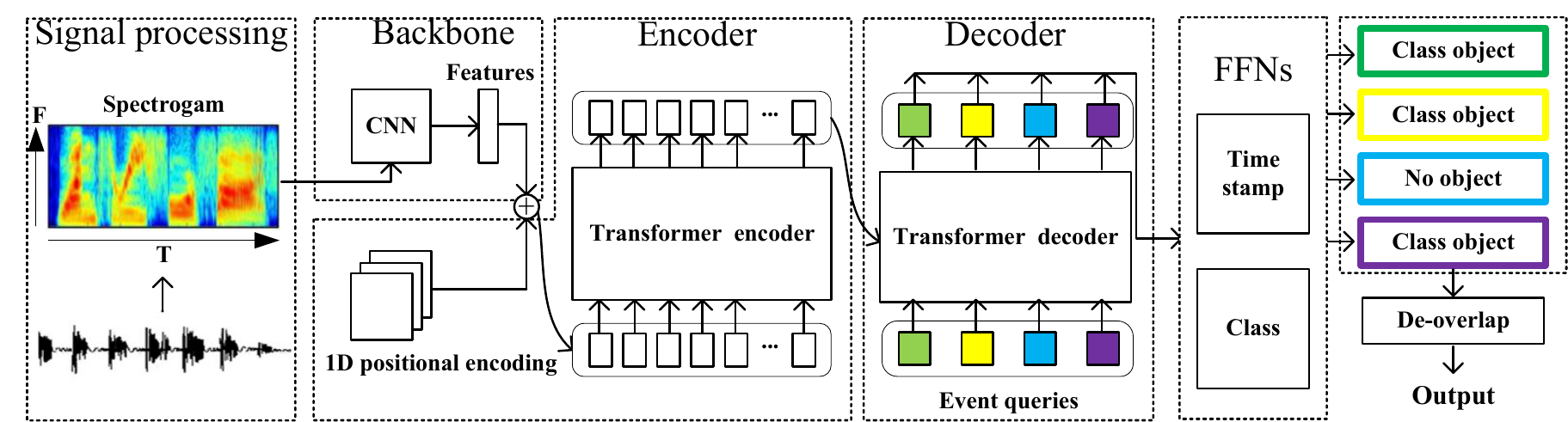}
\caption{Overview of 1D-DETR model.}
\label{1D-DETR}
\end{figure*}
\subsection{Supervised SED}
As mentioned above, polyphonic SED is often formulated as a sequence-to-sequence multi-label classification problem and addressed by classifying each frame and then integrating the frame-level results. The major challenge of polyphonic SED lies in the addictive nature of sound sources, which makes it difficult to find robust features and classifiers. Earlier approaches use the widely applied Mel-Frequency Cepstral Coefficients (MFCC) feature and conventional classifiers from speech recognition. For example, T. Heitto et al.\cite{heittola2013context} use Gaussian Mixture Models (GMMs) to obtain context-dependent frame representation from MFCC and use Hidden Markov Models (HMMs) as the classifier. However, due to the random co-occurrence patterns of sound events and noise interference, such models have poor robustness. To tackle this problem, the spectrogram image feature (SIF) is utilized as input to train CNNs \cite{zhang2015robust}. This CNN-based model shows excellent performance and has been further optimized and adjusted by other researchers \cite{phan2016robust}. However, although the CNN model can capture frame feature with great robustness from the spectrogram, it fails to exploit longer temporal context information, which is essential for sequence related tasks. To address this limitation, CNNs are combined with RNNs, which can model sequential information well, therefore generating CRNNs for SED \cite{cakir2017convolutional}. There are also many follow-up researches on the CRNN architecture \cite{9054433,Adavanne2017spatial}. Recently, with the great success of Transformer and its variants in other fields, Miyazaki et al. \cite{miyazaki2020weakly} propose to replace the RNNs with the encoder of Transformer\cite{vaswani2017attention} or Conformer\cite{gulati2020conformer}.

In order to get event-level results, post-processing is applied to integrate the frame-level predictions to obtain the temporal boundaries, which is crucial to the detection results \cite{cances2019evaluation}. It mainly includes two steps: binarization and median filtering. Some work suggests tuning the binarization thresholds \cite{hou2018semi} and window sizes \cite{DCASE2019ICT} for different event categories on the development set to improve the model's ability to detect events of different durations. However, it only works when the development set and the test set have similar data distribution. What's worse, searching for the best combination of these hyperparameters is quite time-consuming. 

\subsection{SED with weakly-labeled and synthetic data}
Due to the high cost of annotating audios with temporal boundaries of events, it has become a trend to train SED models with weakly-labeled data and synthetic data. 

For weakly supervised SED, which utilizes weakly-labeled data that only indicates the presence of event categories in an audio clip, it is usually approached as a multiple instance learning problem \cite{mil1998maron}, where the frame-level representations are aggregated to obtain a clip-level representation, and then audio tags are predicted to compare with the ground-truth labels \cite{2018Adaptive}. Based on this architecture, many pooling methods have been studied, such as max pooling, average pooling, attention pooling, and so on \cite{2019A, lin2020specialized}. However, whatever methods are adopted to exploit the weak labels, there is always a trade-off between frame-level (event boundary) and clip-level (audio tag) prediction, where the former expects more detailed (frame-level) information, corresponding to a smaller time axis compression scale, while the latter requires more global
information, corresponding to a larger compression scale \cite{2015Fully}. To liberate model from such trade-off, Lin et al. \cite{lin2020guided} propose a teacher-student framework named Guided Learning for SED with weakly-labeled and unlabeled data, where the teacher model focuses on audio tagging, while the student model focuses on the prediction of event boundaries. In addition, Huang et al. \cite{huang2020multi} propose a multi-branch learning method to ensure the encoder can capture more comprehensive feature fit for various subtasks.

Synthetic data is another alternative to train SED models, which is generated by mixing up isolated events with background noise randomly. Synthetic data is often used jointly with real data with strong or weak labels. Initially, synthetic data is used equally to real data, which is obviously not a good choice for there is a large domain gap between synthetic data and real data \cite{Park2019WeaklyLS}. To use synthetic data more effectively, domain adaption is introduced into SED to compensate for the inconsistency of their feature distribution. Adversarial learning and metric learning are two major approaches utilized in existing domain adaptation models for SED \cite{Park2019WeaklyLS, wei2020A-CRNN, huang2020learning}. Different domain adaptation methods may focus on different stages, some aim at data collection condition adaptation \cite{wei2020A-CRNN}, others apply it during model training \cite{cornell:hal-02962911, Park2019WeaklyLS}.

To make soundscape synthesis and augmentation more convenient, Scaper \cite{salamon2017scaper}, an open source library, is developed. Many datasets are generated by it, such as the DESED synthetic soundscapes evaluation set \cite{2018Large}\cite{2020Sound} and the URBAN-SED dataset.

\subsection{DETR for object detection}
Detection Transformer (DETR) \cite{carion2020end} is the first end-to-end model in the field of image object detection, which adopts the encoder-decoder model architecture. DETR exploits the global features of the image by the encoder, learning the possible objects in the image, then employs the decoder to predict the precise position and category. The principle and procedure of DETR are more in line with the top-down human visual attention characteristics, that is, visual attention is usually guided by “goals” in mind, which has been explained by cognitive science in the 
\emph{Preferred Competition Theory} \cite{beck2009top}. Supported by cognitive principles and good performance, DETR has received great attention. Many DETR-related works have been proposed, such as deformable DETR \cite{zhu2020deformable} and unsupervised learning methods for DETR \cite{dai2020up}. It has also been applied to pedestrian detection where detection targets are highly overlapped, and achieves competitive performance, demonstrating its great potential in crowded scenes \cite{2020DETR}.   

However, most work still focuses on image object detection. As mentioned earlier, SED and object detection have much in common. Besides, reports are showing that human auditory attention also follows the top-down manner \cite{2005Some}, which means that the detection principle of DETR is still applicable to SED, while the frame-based model only learns locally.

\section{The 1D-DETR model }
Considering the characteristics of one-dimensional signals, we modify the designs for two-dimensional signals in DETR to obtain 1D-DETR. In this section, we introduce the model architecture and training methods of 1D-DETR.
\subsection{Model Architecture}
As depicted in Figure \ref{1D-DETR}, 1D-DETR consists of three main components: a backbone, an encoder-decoder transformer, and prediction feed-forward networks (FFNs). 1D-DETR represents each sound event as a vector $y_i=(c_i,b_i)$, where $c_i$ is the class label and $b_i=(m_i,l_i)$ denotes the temporal boundary containing normalized event center $m_i$ and duration $l_i$. Given the spectrogram of an audio clip, the backbone and transformer encoder are utilized to extract its feature. The decoder is adopted to generate event representations which are then fed into prediction FFNs to obtain final event detection results. We describe each component in detail below.
\subsubsection{Backbone}
In this paper, we use ResNet-50\cite{7780459} as the backbone, which has achieved excellent performance on audio classification\cite{hershey2017cnn}, confirming its efficiency in audio feature extraction. For an audio clip, its corresponding Mel-spectrogram $X \in \mathbb{R}^{1 \times T_0 \times F_0 }$ will be transformed into a feature map $f\in\mathbb{R}^{C \times T\times F}$, where $T_0$ and $T$ denote the dimensions of the time axis, $F_0$ and $F$ denote the dimensions of the frequency axis, and $C$ denotes the number of channels. Then a $1 \times 1$ convolution is applied to reduce the channels to the number of transformer attention units $d$, resulting in a new feature map $z_0 \in \mathbb{R}^{d \times T \times F}$.
\subsubsection{Positional encoding}
1D-DETR only needs to locate on one axis, so unlike the original DETR, we adopt one-dimensional positional encoding for 1D-DETR, enforcing the model to focus on the time axis. The corresponding formula can be expressed as:
\begin{align}
P_{(t,f,2i)} &=sin(t/10000^{2i/d})\\
P_{(t,f,2i+1)}&=cos(t/10000^{2i/d}) 
\end{align}
where $t, f$ are the two-dimensional coordinates in a Mel-spectrogram,  $i$ is the dimension, and $d$ is the number of transformer attention units. Using these equations, we generate the positional encoding $P\in \mathbb{R}^{d \times T \times F}$ with the same shape as $z_0$. 
\subsubsection{Encoder}
The encoder has a stack of $E$ identical blocks with two sub-layers: a multi-head self-attention mechanism and a feed-forward network. We flatten $z_0$ and $P$ on the time and frequency axis to get a $d \times TF $ feature map and positional encoding to make them meet the input requirements of the encoder. Unlike standard Transformer where positional encoding is only added to the initial input of Transformer, 1D-DETR performs the addition of positional encoding to the input of each layer to enhance its positioning ability.
\subsubsection{Decoder}
The decoder consists of $M$ identical blocks with multi-head self-attention layers, encoder-decoder attention layers, and feed-forward layers. It is employed to generate event representations. Conventional transformers are mainly used to process sequences with strong context dependence, such as text and speech. To capture dependence among outputs, the prediction is made in an incremental token-by-token way with previous output as input. However, for the detection task, the objects are almost independent of each other, so the auto-regressive mechanism would be inappropriate. In 1D-DETR, we model detection as a set prediction problem by using $N$ embeddings as input, where $N$ is a fixed hyperparameter and larger than the typical number of events in an audio. These input embeddings are learned positional encodings, which we call \emph{event queries}, and will be added to the input of each attention layer. This scheme enables 1D-DETR to decode $N$ events in parallel.
\subsubsection{Prediction feed-forward networks (FFNs)}
Prediction FFNs are used to transform the event representations from the decoder into timestamps and class labels. For the timestamp, we represent it as a vector $b_i=(m_i,l_i)$, where $m_i$ and $l_i$ denote normalized event center position and event duration respectively. We use a linear perceptron to obtain the timestamps. For class labels, a simple linear projection layer with a softmax function is used. For each audio, the number of predictions output by 1D-DETR is a fixed number $N$ which is usually larger than the number of ground truth events, hence, an additional ``empty'' class label \text{\o} is added to indicate that certain detection results have detected ``empty'' events.

\subsection{Model Training}

\subsubsection{Matching Method}
A unique matching between predicted and ground truth events is essential for loss computation. In 1D-DETR, we follow the method adopted by DETR\cite{carion2020end} and employ the Hungarian algorithm\cite{kuhn1955hungarian} to obtain an optimal bipartite matching. Let $y_i=(c_i,b_i)$ denote an event, where $c_i$ is the target class label and $b_i \in [0,1]^2 $ denotes the timestamp. Then the matching process can be formulated as searching for a permutation of $N$ elements $\hat{\sigma} \in \mathfrak{S}_{N}$ to get the final matching relation:
\begin{equation}
\hat{\sigma}=\underset{\sigma \in \mathfrak{S}_{N}}{\arg \min } \sum_{i}^{N} \mathcal{L}_{\operatorname{match}}\left(y_{i}, \hat{y}_{\sigma(i)}\right)
\end{equation}
where $y$ is a set of the ground truth events $y_{i}$ with ``empty'' events \text{\o} padded to expand its size to $N$, $\hat{y}=\left\{\hat{y}_{i}\right\}_{i=1}^{N}$ represents the set of $N$ predictions, and $\mathcal{L}_{\operatorname{match}}$ is the corresponding pairwise matching cost, which is defined as $ \hat{p}_{\sigma(i)}\left(c_{i}\right)+ \mathcal{L}_{\text{loc}}\left(b_{i}, b_{\sigma(i)}\right)$ for $c_i \neq \o $ where $\hat{p}_{\sigma(i)}\left(c_{i}\right)$ is the probability of class $c_i$ and $\mathcal{L}_{\text{loc}}$ is the \emph{location loss} which will be defined later. 
\subsubsection{Loss Function}
The loss function is the sum of \emph{location loss} $\mathcal{L}_{\text{loc}}$ and \emph{classification loss} $\mathcal{L}_{\text{c}}$:
\begin{equation}
\mathcal{L}=\mathcal{L}_{\text{loc}}+\mathcal{L}_{\text{c}} \label{eq4}
\end{equation}
The \emph{location loss} is obtained by computing the L1 norm between the target and predicted location vector. In order to speed up the convergence during training, IOU loss is further applied:
\begin{align}
\mathcal{L}_{\text{loc}}&=\sum_{i}^{N}\mathbbm{1}_{\left\{c_{i} \neq \emptyset \right\}}\mathcal{L}_{\text{loc}}(b_i, \hat{b}_{\hat{\sigma}(i)}) \\ \nonumber
&=\sum_{i}^{N}\mathbbm{1}_{\left\{c_{i} \neq \emptyset \right\}} \left(\lambda_{\text{IOU}}\mathcal{L}_{\text{IOU}}(b_i,\hat{b}_{\hat{\sigma}(i)})+\lambda_{\text{L1}}\parallel b_i-\hat{b}_{\hat{\sigma}(i)} \parallel_1 \right) 
\end{align}
where $\lambda_{\text{IOU}}, \lambda_{\text{L1}}\in \mathbb{R}$ are hyperparameters, $\hat{\sigma}$ is the optimal assignment given by the matching process, and $N$ is the number of predictions. The \emph{classification loss} is the cross-entropy between the labels and the predictions:
\begin{equation}
\mathcal{L}_{\text{c}}=\sum_{i=1}^N -\operatorname{log}\hat{p}_{\hat{\sigma}(i)}(c_i) \label{eq6}
\end{equation}
The prediction loss of each decoder block is calculated to assist the model training in 1D-DETR. To sum up, the loss function of the entire network can be expressed as:
\begin{equation}
\mathcal{L}_{\text{total}}=\sum_{m=1}^{M}\mathcal{L}^\text{m}    
\end{equation}
where $M$ represents the number of decoder blocks.

\section{Sound Event Detention Transformer}
\begin{figure}[!t]
\centering
\includegraphics[width=3.5in]{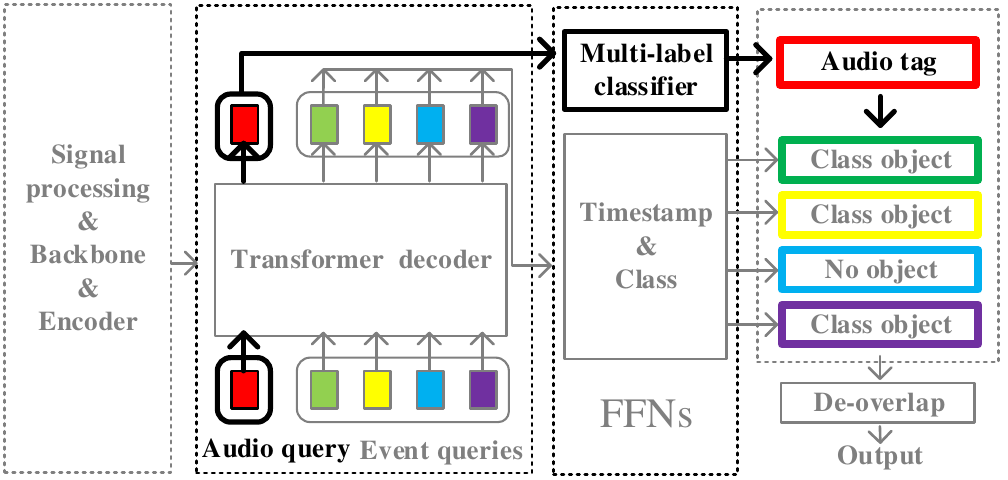}
\caption{Overview of the audio query branch of SEDT, drawn with the bold black line, including Audio query and Multi-label classifier.}
\label{SEDT}
\end{figure}
When applying 1D-DETR to SED, we found many predictions with accurate positions but wrong class labels. We conjecture the poor classification performance of 1D-DETR is mainly due to two reasons: the insufficient category information extracted by the model and the one-to-one matching principle adopted during training. To deal with these two problems, on the basis of 1D-DETR, we propose an audio query branch and one-to-many matching strategy, obtaining Sound Event Detection Transformer (SEDT).

\subsection{Audio query branch}
As mentioned in \cite{carion2020end}, DETR may wrongly classify a certain number of accurately positioned predictions as the ``empty'' class. To optimize the performance of DETR, these predictions are overridden with the second-highest scoring classes in the work of \cite{carion2020end} . But it is not a good choice, because sometimes DETR does detect accurate ``empty'' events. We explored the cause of this problem and speculated that the poor classification performance may be due to the decreased category information extraction ability of the model caused by the positioning subtask and the introduction of the ``empty'' class. To verify this conjecture, we compared the backbone network trained within the 1D-DETR targeting SED with the same network trained individually only targeting event classification (audio tagging) by connecting both trained networks to a classifier and measuring the audio tagging performance of them (the details will be described in Section \ref{ablation}).  We found that the backbone trained individually performed better than the backbone trained within 1D-DETR, indicating that 1D-DETR might lose a lot of category information in the backbone feature extraction stage.

Based on the above observation, we introduce an audio query branch to 1D-DETR. On one hand, the added clip-level classification branch can enhance the category information in the model.  On the other hand, we employ the audio tag to assist event-level decision-making, which is more reasonable compared with the method adopted in DETR which replaces the ``empty'' class with the second-highest scoring class.
\subsubsection{Audio query}
We hope the model can better recognize events without reducing the positioning accuracy, thereby improving the event-based metric. The key is to guarantee there is enough category information during decoding. 
We introduce an audio query, which is also a learned positional embedding, similarly to event queries. The audio query and event queries are sent to the decoder together. We assume the output corresponding to audio query aggregates the whole sequence information and feed it into a multi-label classifier which is a fully connected layer with Sigmoid function to get the predicted weak label $\boldsymbol{l}_{\text{TAG}}$ as illustrated in Figure \ref{SEDT}. There would be an extra \emph{audio tagging loss} for SEDT, which is calculated as the binary cross-entropy (BCE) between $\boldsymbol{l}_{\text{TAG}}$ and the weak label $\boldsymbol{y}_{\text{TAG}}$:
\begin{equation}
\mathcal{L}_{\operatorname{at}}=\operatorname{BCE}(\boldsymbol{l}_{\text{TAG}},\boldsymbol{y}_{\text{TAG}})
\end{equation}
\subsubsection{Event-level classification fusion}
\begin{figure}[!t]
\centering
\subfloat[]
{\includegraphics[width=3.5in]{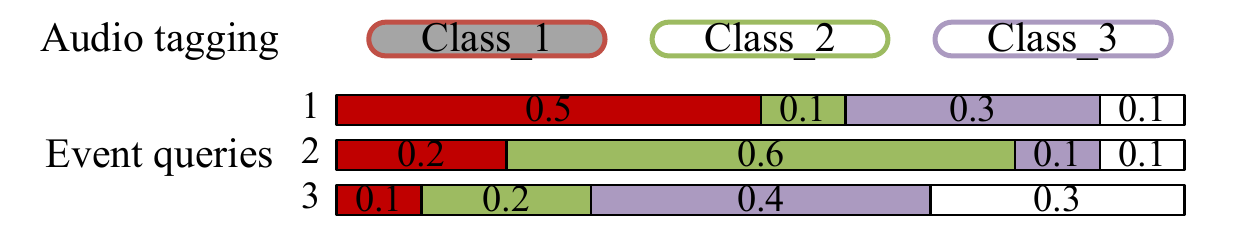}%
\label{a}}
\hfil
\subfloat[]
{\includegraphics[width=3.5in]{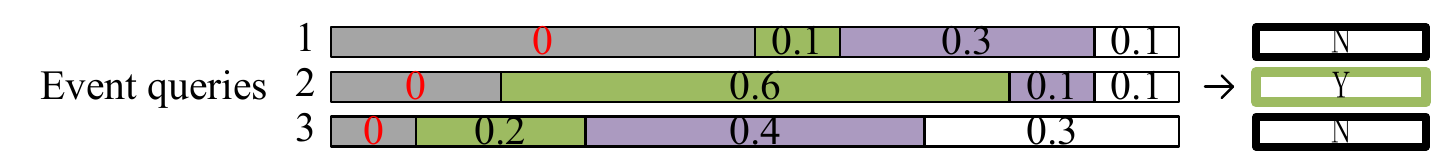}%
\label{b}}
\hfil
\subfloat[]
{\includegraphics[width=3.5in]{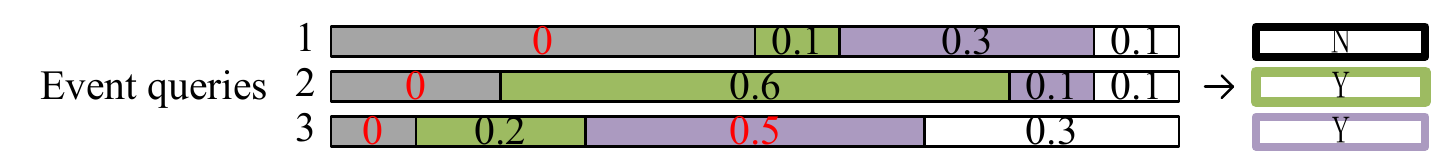}%
\label{c}}
\hfil
\subfloat[]
{\includegraphics[width=3.5in]{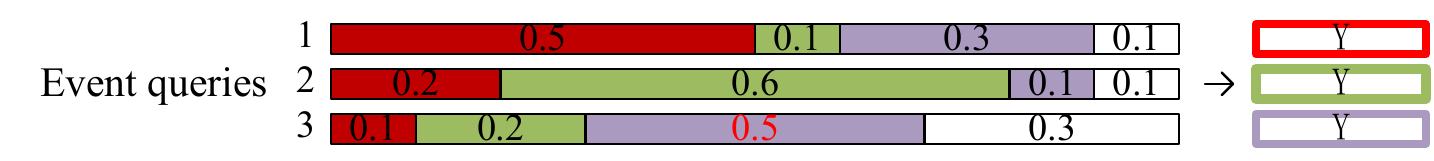}%
\label{d}}
\caption{Illustration of three event-level classification fusion strategies with three event categories and three event queries. The audio tagging results are represented by rounded rectangles, where different borderline colors indicate different event categories and gray filling indicates the corresponding categories are judged as inactive, while white filling as active. For event queries, the illustration shows the probability distribution of all event classes (including the ``empty'' class). Figure (a) is the original output, (b)-(d) give the probability distribution and the final output under corresponding strategies, where ``Y'' means the predicted event is regarded as the final output, while ``N'' means not.}
\label{strategy}
\end{figure}
Compared with SED, audio tagging is a relatively simple task, a model for which can achieve high recognition accuracy. Thus we expect the tagging results can further assist to improve SED performances, and we propose three strategies to fuse the event-level and clip-level classification results:

\textbf{Strategy 1:} Delete the detected events with class labels which are predicted by audio tagging as inactive. As depicted in Figure \ref{strategy} (b), for all event queries, the probabilities of class $1$ are set to $0$, so that no predictions for class $1$ will be retained in the final output.

\textbf{Strategy 2:} On the basis of strategy $1$, ensure that there exist detected events with class labels that are predicted as active by audio tagging. Specifically, find the detected event with the largest class probability for each class that is predicted as active by audio tagging and set its corresponding class probability as the threshold if its original probability is smaller than the threshold. In Figure \ref{strategy} (c), Class $3$ is predicted as active by audio tagging. The largest probability of class $3$ lies in event query $3$ but is smaller than the threshold, which we set as $0.5$. Hence, we modify it as $0.5$ to ensure the occurrence of class $3$ in predictions.

\textbf{Strategy 3:} Ensure there are detected events with class labels predicted as active by audio tagging, and the specific implementation is the same as strategy $2$. Compared with strategy $2$, strategy $3$ will not set the probability of class $1$ as $0$.

Let $C_{\operatorname{audio\_{tagging}}}$ and $C_{\operatorname{SED}}$ denote the active class set predicted by audio tagging and SED respectively, then the class set after fusion can be correspondingly expressed as:

\textbf{Strategy 1:} $C=C_{\text{audio\_{tagging}}}  \bigcap  C_{\text{SED}}$

\textbf{Strategy 2:} $C=C_{\text{audio\_{tagging}}}$

\textbf{Strategy 3:} $C=C_{\text{audio\_{tagging}}} \bigcup C_{\text{SED}}$

\subsection{One-to-many matching strategy}
\begin{algorithm}[h]
  \caption{Model training with one-to-many matching}
  \label{code:alg1}
  \begin{algorithmic}[1]
    \Require $\eta = $ the initial learning rate
    \Require $x^k =$ the $k$th training sample
    \Require $S_\theta(x^k) = $ set of predictions of $x^k$
    \Require ${y}^k = $ set of ground-truth events of $x^k$
    \Require $G = $ the number of ground-truth events in the audio
    \Require $N = $ the number of predictions for each input
    \Require $\mathcal{L}_{\text{match}}$ = pair-wise matching loss function
    \Require $\mathcal{L}_{\text{loc}}$ = location loss function
    \Require $\mathcal{L} = $ loss function
    \Require $\operatorname{Sample}_{\frac{\alpha G}{N}}$ = randomly sample function at the ratio of $\frac{\alpha G}{N}$;
    \Ensure
    $\theta$
    \For{$epoch = 1 \to epoches$}
        \State $ \eta \gets 0.1^{\lfloor \frac{epoch}{epochs_{drop}}\rfloor} \times \eta $
        \For{each batch $B$}
            \State $\hat{y}^k \gets S_\theta(x^k)$
            \State $\hat{\sigma}^k=\underset{\sigma \in \mathfrak{S}_{N}}{\arg \min } \sum_{i}^{N} \mathcal{L}_{\operatorname{match}}\left(y_{i}^k, \hat{y}_{\sigma(i)}^k\right)$
            \If {$epoch < epoches_{lr} $} \Comment{the learning stage}
                \State $loss \gets \frac{1}{|B|}\sum_{k}\sum_{i}^{N}\mathcal{L}(y_i^k,\hat{y}_{\hat{\sigma}^k(i)})$
            \Else \Comment{the fine-tuning stage}
                \State $I_{j}^k \gets \mathop{\arg\min}_{i} \mathcal{L}_{\text{loc}}(y_i^k, \hat{y}_j^k)$
                \State $D_j^k \gets \min_{i} \mathcal{L}_{\text{loc}}(y_i^k, \hat{y}_j^k)$
                \State $\Sigma_{\epsilon}^k \gets \left\{ j|D_j^k<\epsilon, j \in \left\{\hat{\sigma}^k(\o)\right\} \right\}$
                \State $\hat{\Sigma}_{\epsilon}^k \gets \operatorname{Sample}_{\frac{\alpha G}{N}}(\Sigma_{\epsilon}^k)$
                \State $\hat{\Sigma}_{\epsilon}^k(i) \gets \left\{j|I_j^k==i, j\in \hat{\Sigma}_{\epsilon}^k\right\}$
                \State $\hat{\Sigma}^k(i) \gets \hat{\Sigma}_{\epsilon}^k(i) \cup \left\{\hat{\sigma}^k(i)\right\}$
                \State $loss \gets \frac{1}{|B|}\sum_{k}\sum_{i}^{N}\sum_{m \in \hat{\Sigma}^k(i)}\mathcal{L}(y_i^k,\hat{y}_m^k)$
            \EndIf
            \State update $\theta$ \Comment{update network parameters}
        \EndFor
    \EndFor
  \end{algorithmic}
\end{algorithm}
As mentioned earlier, the number of predictions $N$ by 1D-DETR is usually larger than the number of ground-truth events, resulting in a considerable proportion of the predictions with accurate boundaries being matched to ``empty'' events during training, so the model is inclined to judge some events as ``empty'' class during testing even though they locate correctly. The root of the above problem is that the Hungarian algorithm only permits one-to-one matching. So a natural choice is to relax the one-to-one restriction and allow one-to-many matching. In this section, we introduce the training process which adopts the one-to-many matching strategy, as shown in Algorithm \ref{code:alg1}.

The one-to-many matching strategy not only considers the matching obtained by the Hungarian algorithm but also introduces matching whose \emph{location cost} is less than a preset value $\epsilon$. The \emph{location cost} of a prediction is defined as the smallest \emph{location loss} between itself and all targets. In practice, the following strategy is employed to obtain the one-to-many matching pairs: First, the Hungarian algorithm is applied to get one-to-one matching $\hat{\sigma}$ as DETR does. Then, the \emph{location cost} of predictions that are matched with ``empty'' events in Hungarian matching is calculated. After that, we can get a set ${\Sigma}_{\epsilon}$ of predictions with \emph{location cost} smaller than a preset value $\epsilon$. Finally, predictions from ${\Sigma}_{\epsilon}$ are randomly selected at the ratio of $\frac{\alpha G}{N}$ to obtain $\hat{\Sigma}_{\epsilon}$, where $\alpha$ is a hyperparameter, $G$ is the number of ground-truth events and $N$ is the number of predictions for each audio. The selected predictions are matched with targets with which the \emph{location cost} are obtained. Under this strategy, a ground truth event may match multiple predictions, thereby reducing the ``empty'' class matching. 

In \cite{carion2020end}, DETR learns different specialization for each query slot, that is, each slot focuses on targets of different sizes and positions. Unfortunately, we observe that the one-to-many matching strategy can undermine this specialization. We speculate this is because, under this strategy, some slots are optimized to detect the same target, leading to homogeneity. To overcome this shortcoming, in our work, the training of SEDT is carried out in two stages. One adopts the one-to-one matching strategy to learn the slot with specialization, referred to as the \emph{learning stage}. The other adopts the one-to-many matching strategy for the alleviation of ``empty'' class predictions, referred to as the \emph{fine-tuning stage}.

The fine-tuning stage makes one target corresponds to multiple predictions, leading to a large overlapping between predictions. To remove redundant predictions, we perform de-overlapping on the output of SEDT.
De-overlapping is only performed for predictions with the same class label. Class probability is applied as the retention metric, that is, only the result with the highest class probability in the overlapped predictions is retained.
\subsection{weakly-supervised SEDT}
SEDT is trained by minimizing both clip-level and event-level loss function. However, weakly-labeled data only indicates the presence of event categories in a clip, which means the event-related modules of SEDT, that is, prediction FFNs for final timestamp and class label output, will not be optimized. To equip SEDT with the ability of weakly supervised learning, we perform max pooling operation on the event-level class probability vectors to obtain a clip-level class probability vector.  Then we calculate the binary cross-entropy (BCE) between this pooling probability vector $\boldsymbol{p}_{\text{TAG}}$ and the weak label $\boldsymbol{y}_{\text{TAG}}$ to obtain \emph{pooling audio tagging loss}, which can be formulated as
\begin{equation}
\mathcal{L}_{\operatorname{at}}^{\operatorname{p}}=\operatorname{BCE}(\boldsymbol{p}_{\text{TAG}},\boldsymbol{y}_{\text{TAG}})
\end{equation}
Obviously, \emph{pooling audio tagging loss} still fails to update the linear perceptron for timestamp prediction. We assort to synthetic data to train this module. In this way, we can train SEDT by real weakly-labeled data and synthetic data, which is a kind of weakly supervised learning.
\subsection{Loss function}
\subsubsection{Loss for strongly-labeled data}
For training audios with annotation of temporal boundaries of sound events, the SEDT is trained by minimizing the linear combination of \emph{audio tagging loss} and the basic loss of 1D-DETR, including \emph{location loss} and \emph{classification loss}:
\begin{align}
    \mathcal{L}_{\text{strong}}&= \mathcal{L}_{\text{1D-DETR}}+\lambda_{at}\mathcal{L}_{\text{at}}\\ \nonumber
    &=\sum_{m=1}^{M}\left(\mathcal{L}_\text{loc}^m+\mathcal{L}_{\text{c}}^m\right)+\lambda_{at}\mathcal{L}_{\text{at}}
\end{align}
where $\lambda_{at}$ is a hyperparameter.
\subsubsection{Loss for weakly-labeled data}
The predicted weak labels by audio query and pooling operation allow the model to learn the distribution characteristics of weakly-labeled audios. For weakly-labeled data, only \emph{audio tagging loss} and \emph{pooling audio tagging loss}  are computed: 
\begin{align}
    \mathcal{L}_{\text{weak}}&=\lambda_\text{at}\mathcal{L}_{\text{at}}+\lambda_\text{at}^p\mathcal{L}_{\text{at}}^p
\end{align}
We can train a SEDT with real weakly-labeled data and synthetic strongly-labeled data. The corresponding loss function can be formulated as
\begin{align}
    \mathcal{L} & = \mathcal{L}_{\text{strong}}^{\text{synthetic}} + \mathcal{L}_{\text{weak}}^{\text{real}}
\end{align}
With the knowledge of real data learned by audio query and pooling operation, it can be used to detect events in real audios.
\section{Experiments}
\subsection{Experimental setup}
\subsubsection{Dataset}
We perform experiments on the URBAN-SED and DCASE2019 Task4 datasets. URBAN-SED \cite{2014A} is a dataset of synthesized audio clips generated by Scaper \cite{salamon2017scaper}, and the data is with annotation of temporal boundaries of sound events (strongly-labeled). It contains 10 different categories of sound events and is divided into a training set (6000 clips), a validation set (2000 clips), and a test set (2000 clips). The DCASE2019 Task4 dataset also has 10 sound event categories and is divided into a training set and a validation set. The training set consists of three subsets, i. e., a weakly-labeled dataset of real-world recordings (1578 clips), a synthetic dataset with strong labels (2045 clips) and an unlabeled dataset of real-world recordings (14412 clips). The validation data (1168 clips) are real recordings with strong labels. The DCASE2019 Task4 dataset is the same as the DCASE2021 Task4 dataset except for the synthetic subset. In our experiment, the unlabeled subset will not be used. We extract 64-dimensional log-Mel spectrograms from all the data as input features. 
\subsubsection{Model}
For SEDT, ResNet-50 is adopted as the backbone model, and Transformer with 3 encoder and 3 decoder layers is employed. The number of event queries is set as $10$ for the URBAN-SED dataset and $20$ for the DCASE2019 Task 4 dataset. We use the following rules to name the SEDT models in our experiments: the suffix ``AQ'' denotes models with audio query, ``FT'' denotes models after fine-tuning with one-to-many matching, and ``P$\emph{m}$'' denotes event-level classification fusion strategy $\emph{m}$ is applied.

As for the baselines, we follow the model architecture in \cite{miyazaki2020weakly} to build the CRNN-based \cite{turpault2019sound} and the Transformer-based model \cite{miyazaki2020weakly}, which is referred to as ``CTrans'' in this paper. And on the DCASE2019 Task4 dataset, the synthetic data is used as strongly-label data as in \cite{miyazaki2020weakly}. For CRNN and CTrans, the frame-level predictions are smoothed by a median filter with a fixed window size of $0.45s$ for different event categories. To demonstrate that SEDT can learn the duration of different event categories, we also give the results of CRNN and CTrans post-processed by adaptive median filtering \cite{DCASE2019ICT}, that is, the window size for each event category is optimized on the validation dataset. We denote them as CRNN-CWin and CTrans-CWin, respectively. The class probability threshold is fixed as 0.5 to determine sound events for both SEDT and baseline models. 


\subsubsection{Training}
The models are trained with a batch of 128 clips with the AdamW optimizer \cite{loshchilov2017decoupled}. In the learning stage, the model is trained for $400$ epochs with an initial learning rate of $10^{-4}$. Then the best model obtained in the learning stage is fine-tuned for another $100$ epochs, with initial learning rate of $10^{-5}$. The hyperparameters of loss function are set as follow: $\lambda_\text{at}=\lambda_\text{at}^p=0.25$ for the DCASE2019 Task4 dataset, $\lambda_\text{at}=3$ for the URBAN-SED dataset, and $\lambda_{\text{IOU}}=2, \lambda_\text{L1}=5$ for both datasets. As for the fine-tuning stage, the setup of $\epsilon=1, \alpha=1$ is used.
\subsubsection{Evaluation}
To evaluate the performance of models, we compute the event-based measure (a 200ms collar on onsets and a $200ms/20\%$ of the events length collar on offsets) and the segment-based measure (the length of a segment is 1s) by the sed\_eval package \cite{mesaros2016metrics}. We also compute the clip-level macro F1 score to measure the performance of models on audio tagging. For the SEDT models, the audio tagging results are given by the audio query branch.
\subsection{Results of 1D-DETR and SEDT}
\subsubsection{Results on the URBAN-SED dataset}
\begin{table}[t!]
    \centering
    \caption{Performance on URBAN-SED test set}
    \begin{tabular}{cccc}
    \toprule
        Model & Eb$[\%]$ & Sb$[\%]$ & At$[\%]$\\
        \midrule
         CRNN\cite{martin2019sound} & - & $64.30$ & - \\
         CRNN & $35.26$ & $65.75$ & $74.64$\\
         CRNN-CWin & $36.75$ & $65.74$ & $74.19$ \\
         \midrule
         CTrans & $31.33$ & $64.51$ & $74.67$ \\
         CTrans-CWin & $34.36$ & $64.73$ & $74.05$ \\
         \midrule
         1D-DETR & $32.71$ & $60.64$ & $70.90$ \\
         SEDT-AQ-FT-P1 & $\bm{37.27}$ & $65.21$ & $74.37$ \\
         SEDT-AQ-FT-P2 & $36.61$ & $65.53$ & $75.12$ \\
         SEDT-AQ-FT-P3 & $36.36$ & $\bm{65.77}$ & $\bm{75.61}$\\
         \bottomrule
    \end{tabular}
    \label{tab:urban}
\end{table}
In the published work \cite{martin2019sound}, CRNN achieved SOTA results on the URBAN-SED dataset with a segment-based metric of $64.30\%$, while the event-based metric and audio tagging results are not given. So we train the DCASE2020 Task4 baseline model, a well-designed CRNN model, as one of our baselines. The CTrans model follows the architecture proposed by \cite{miyazaki2020weakly}, and the hyperparameters of Transformer encoder are consistent with that of SEDT.

As shown in Table \ref{tab:urban}, where Eb, Sb and At denote Event-based, Segment-based, and Audio tagging macro F1 respectively, the SEDT models obtain results competitive to the baselines, which shows the effectiveness of our event-based end-to-end SED framework. SEDT-AQ-FT-P1 achieves the best event-based F1 score of $37.27\%$, and SEDT-AQ-FT-P3 achieves the best segment-based and audio tagging F1 score. Although class-adaptive median filtering can bring benefits, CRNN-CWin is still inferior to SEDT-AQ-FT-P1. The results demonstrate that the proposed model can take both local and global information of the audio into account. The positioning accuracy required by event-based, segment-based, and audio-based metrics is gradually reduced, and strategy 1, 2, and 3 have gradually relaxed requirements for outputs. Thus they can cater to different metrics and achieve the best F1 score on the corresponding metric.

From table \ref{tab:urban}, we can also observe that the CTrans model performs worse than the CRNN model, but it gets more improvement from class-adaptive median filtering, from $31.33\%$ to $34.36\%$, which is consistent with the experiment results in \cite{miyazaki2020weakly}. The CTrans-CWin does not outperform the CRNN-CWin model, which may be due to that the URBAN-SED dataset is a synthetic dataset, where the duration of different event categories does not vary significantly thus the class-adaptive median filtering can not bring much extra event-related information. 
\begin{table}[t!]
    \centering
    \caption{Performance on DCASE2019 Task4 validation set}
    \begin{tabular}{cccc}
    \toprule
         Model & Eb$[\%]$ & Sb$[\%]$ & At$[\%]$\\
         \midrule
         CRNN\cite{miyazaki2020weakly} & $23.61$ & $61.00$ & -\\
         CRNN-CWin & $28.59$ & $61.48$ & $65.13$ \\
         \midrule
         CTrans\cite{miyazaki2020weakly} & $21.31$ & $62.44$ & - \\
         CTrans-CWin & $30.72$ & $\bm{63.91}$ & $68.70$ \\
         \midrule
         SEDT-AQ-P1 & $\bm{31.64}$ & $61.18$ & $69.11$ \\
         SEDT-AQ-P2 & $31.15$ & $61.03$ & $69.09$ \\
         SEDT-AQ-P3 & $29.52$ & $59.94$ & $69.32$ \\
         \midrule
         SEDT-AQ-FT-P1 & $29.29$ & $62.71$ & $\bm{70.97}$\\
         SEDT-AQ-FT-P2 & $29.19$ & $62.68$ & $70.93$\\
         SEDT-AQ-FT-P3 & $28.04$ &	$61.52$ & $70.97$\\
         \bottomrule
    \end{tabular}
    \label{tab:dcase}
\end{table}
\begin{table*}[t!]
    \centering
     \caption{Results of ablation experiments on URBAN-SED test set}
    \begin{tabular}{cccccccc}
    \toprule
    \multirow{2}{*}{Model}& \multicolumn{3}{c}{Event-based $[\%]$}&\multicolumn{3}{c}{Segment-based$[\%]$}&Audio tagging$[\%]$\\
     &$\rm{F1}$&$\rm{P}$&$\rm{R}$&$\rm{F1}$&$\rm{P}$&$\rm{R}$&$\rm{F1}$ \\
    \midrule
     1D-DETR(2D-pos) & $31.99$ & $38.39$ &$27.86$ & $59.63$ & $71.20$ & $51.99$ &	$69.98$\\
     1D-DETR(1D-pos) & $32.71$ & $38.43$ & $28.86$ & $60.64$ & $70.89$ & $53.55$ & $70.90$\\
     \midrule
     SEDT-AQ & $33.62$ & $39.86$ & $29.57$ & $61.02$ &	$72.36$ & $53.53$ &	$71.17$\\
     SEDT-AQ-P1 & $34.92$ & $42.64$ & $30.15$	& $62.08$ & $\bm{75.65}$ & $53.49$ & $71.76$\\
     SEDT-AQ-P2 & $35.00$ & $40.29$ & $31.46$ & $64.32$ & $73.72$ & $57.85$ & $74.67$ \\
     SEDT-AQ-P3 & $34.80$ & $39.16$ & $31.70$ & $63.94$ & $71.75$ & $58.27$ & $74.82$ \\
     \midrule
     SEDT-AQ-FT &$36.72$ & $41.63$ & $33.30$ & $65.17$ & $73.19$ & $59.38$ & $74.75$\\
     SEDT-AQ-FT-P1 & $\bm{37.27}$ & $\bm{43.32}$ & $33.21$ & $65.21$ & $\bm{74.82}$ & $58.46$ & $74.37$ \\
     SEDT-AQ-FT-P2 & $36.61$ & $41.36$ & $33.25$ & $65.53$ & $72.91$ & $60.13$ & $75.12$ \\
     SEDT-AQ-FT-P3 & $36.36$ &	$39.95$ & $\bm{33.79}$ & $\bm{65.77}$ & $71.28$ &	$\bm{61.70}$ & $\bm{75.61}$\\
    \bottomrule
    \end{tabular}
    \label{tab:ablation}
\end{table*}
\begin{figure*}[!t]
\centering
\subfloat[Raw]{\includegraphics[width=1.8in]{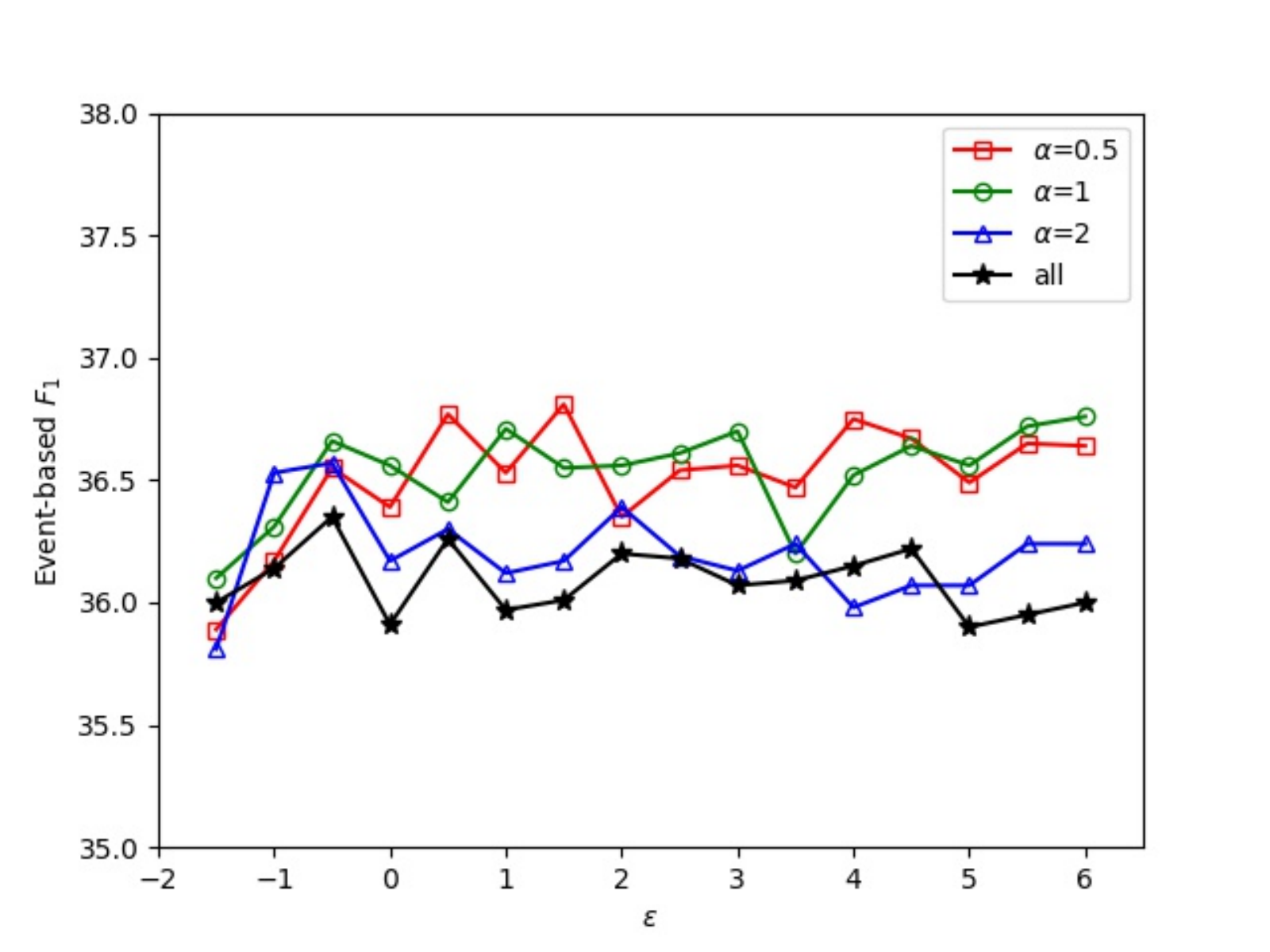}%
\label{sa}}
\subfloat[Strategy 1]{\includegraphics[width=1.8in]{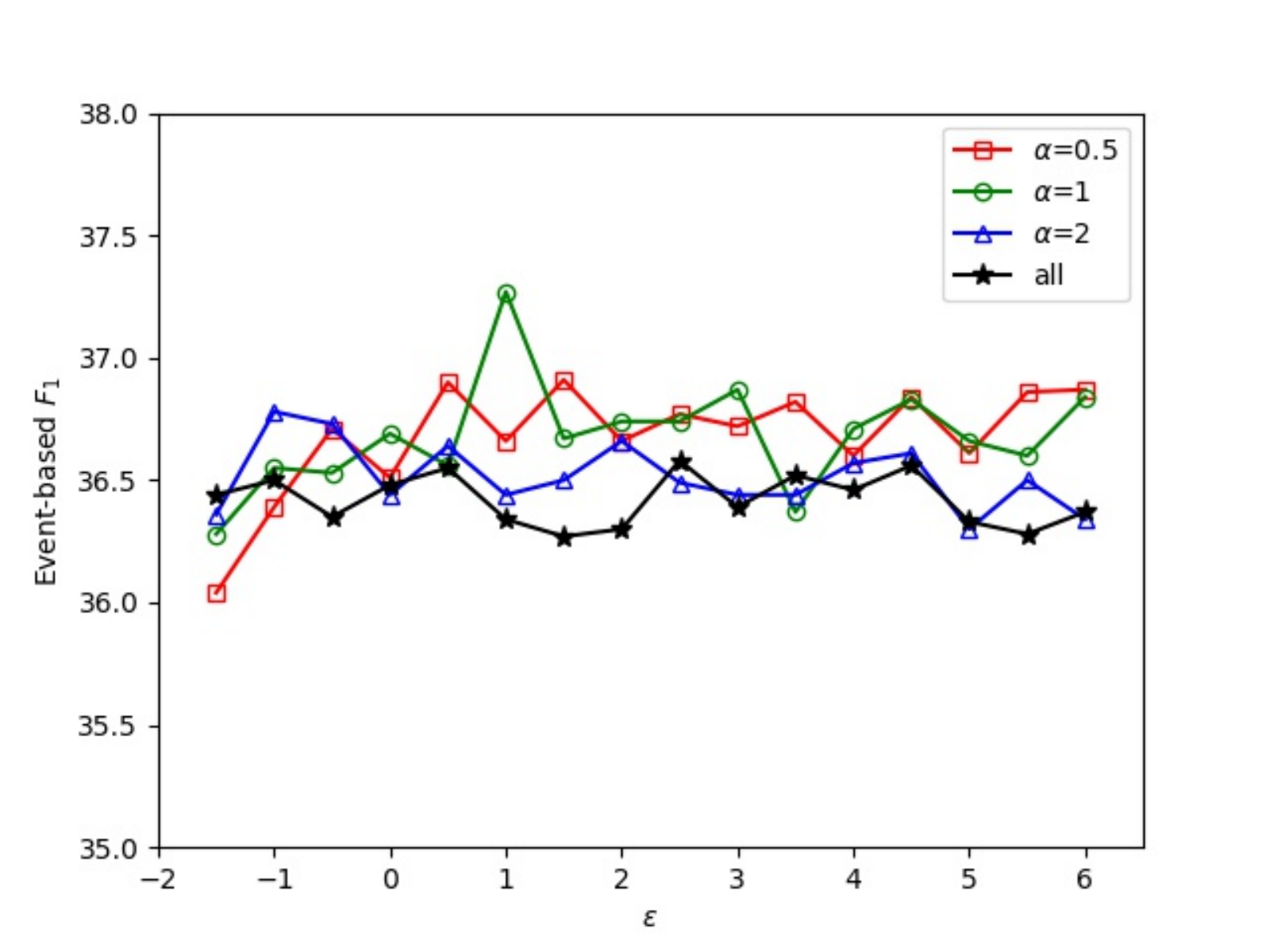}%
\label{sb}}
\subfloat[Strategy 2]{\includegraphics[width=1.8in]{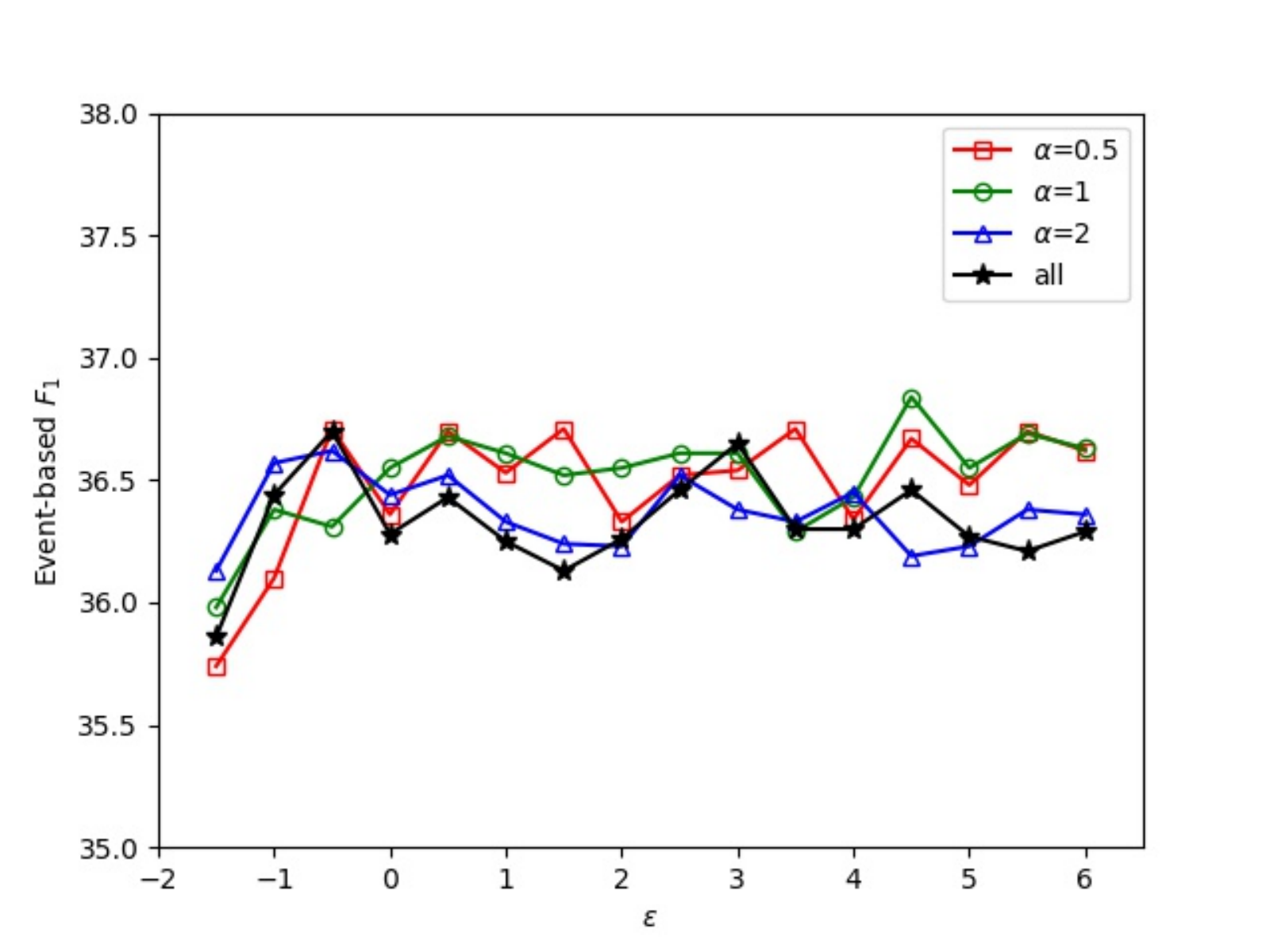}%
\label{sc}}
\subfloat[Strategy 3]{\includegraphics[width=1.8in]{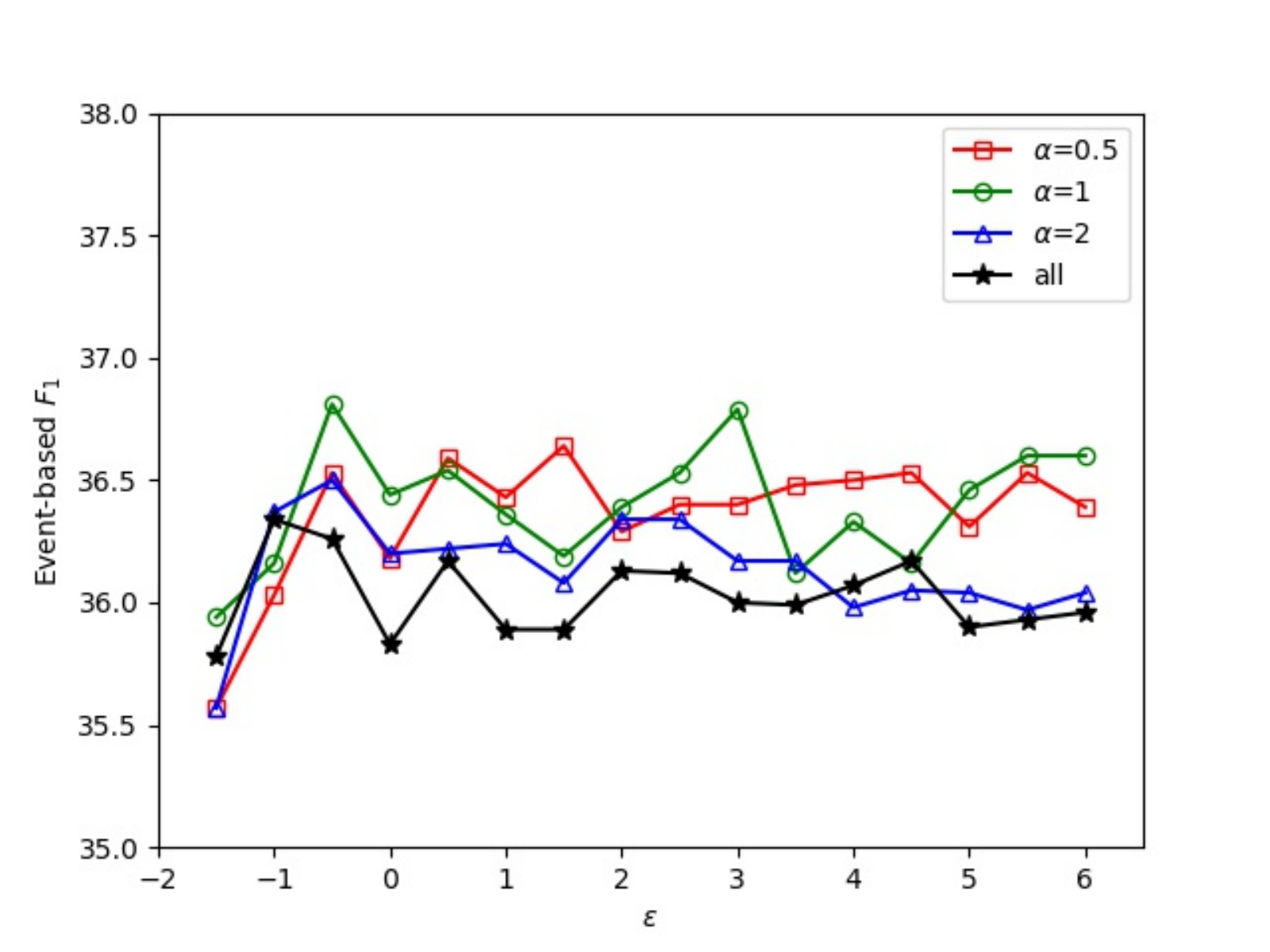}%
\label{sd}}
\caption{The impact of $\epsilon$ and $\alpha$ on the performance of models with different strategies applied.}
\label{parameter}
\end{figure*}
\subsubsection{Results on the DCASE2019 Task4 dataset}
On this dataset, SEDT can only learn to predict event boundaries from synthetic data, while the real data can only be learned through the audio tagging task. We measure the performance of SEDT on the validation set, which is composed of real-recording clips. The results are given in Table \ref{tab:dcase}, where Eb, Sb, and At denote Event-based, Segment-based, and Audio tagging macro F1 respectively.

 Results show that for the main evaluation metric in SED, i.e. event-based metric, SEDT-AQ-P1 has a significant advantage over the CRNN and CTrans. Compared with the baselines with a fixed threshold and the class-adaptive window size, SEDT-AQ-P1 still outperforms CRNN-CWin and  CTrans-CWin by $3.05\%$ and $0.92\%$ respectively. Since the test set is not public, we optimize the window size and evaluate the model on the same dataset, i.e. the validation set. Hence, the class-adaptive window size will bring much more improvement on the DCASE2019 Task4 dataset than the URBAN-SED dataset. Considering that SEDT does not exploit information from the validation set, we argue that SEDT has greater potential than the baselines. For audio tagging, the SEDT models also outperform the baselines. For the segment-based metric, the CTrans-Cwin achieves the best F1 score. This may be because that the SEDT is trained with the objective of accurate event temporal boundaries, tending to reduce predictions with longer duration than the ground-truth, while the frame-based models have more relaxed localization requirment.

Table \ref{tab:dcase} shows that after fine-tuning, the event-based performance of the SEDT models have decreased slightly, which is contrary to the results on the URBAN-SED dataset. The one-to-many matching used in fine-tuning sacrifices the specialization of event query slots to make the model fit the training set better. When the distribution of the training set and the test set are the same (URBAN-SED), this strategy can work, but when the two sets are quite different (the training data of the DCASE2019 dataset contains synthetic data while the data in the validation set are all real recordings), it may cause performance degradation instead. However, the audio tagging performance increases after fine-tuning, and SEDT-AQ-FT-F1 performs best among all models. The segment-based metric also improves due to the excellent performance of audio tagging. This shows that the audio tagging task is less affected by the difference between the training and the test data and can benefit from the one-to-many matching strategy.
\subsection{Ablations}
\label{ablation}
In this paper, we modified the two-dimensional positional encoding of DETR into one-dimensional positional encoding, obtaining the 1D-DETR model suitable for one-dimensional signals. And based on 1D-DETR, the audio query and fine-tuning method based on the one-to-many matching strategy are added to form the Sound Event Detection Transformer (SEDT) model. To verify the effectiveness of these modifications, we perform ablation experiments on the URBAN-SED dataset.
\subsubsection{Positional encoding}
As shown in Table \ref{tab:ablation}, on all of the three metrics, 1D-DETR with two-dimensional positional encoding, denoted as 1D-DETR (2D-pos), lags 1D-DETR with one-dimensional positional encoding, denoted as 1D-DETR (1D-pos), by about $1\%$, verifying that the two-dimensional positional encoding brings redundant information, which misleads the model and worsens its detection performance.
\subsubsection{Audio query}
We investigate the effects of audio query of the SEDT model. The experimental results are shown in Table \ref{tab:ablation}. SEDT-AQ outperforms 1D-DETR by $0.91\%$ and $0.38\%$ in event-based and segment-based macro F1 respectively. We guess that this improvement is due to that the audio query guides the model to learn more category information during the feature extraction stage of the backbone, which will be sent to the decoder, thus improving its classification performance. To verify our conjecture, we fix the parameters of the backbone of the trained 1D-DETR and the SEDT-AQ and use them as feature extractors to construct two audio tagging models by cascading classifiers after them respectively. The macro F1 score is computed to evaluate their performance on audio tagging. We also give the audio tagging results when the backbone and the classifier are trained jointly to evaluate the network structure. The results are shown in Table \ref{tab:backbone}. The model with SEDT-AQ backbone achieves an average macro F1 score of $75.44\%$ and outperforms the model with 1D-DETR backbone, suggesting that SEDT with audio query learns more category information than 1D-DETR. But it still lags behind the model with a trainable backbone by $1.80\%$. This is because the model needs to weigh between classification and positioning.

We continue to compare different event-level classification fusion strategies. We note that strategy 1 achieves the best $\rm{P}$ but the worst $\rm{R}$ on both event-based and segment-based metrics, while on the contrary, strategy 3 achieves the worst $\rm{P}$ but the best $\rm{R}$. This is reasonable because strategy 3 uses more predictions as the final output, will cover more correct predictions to improve $\rm{R}$ and introduce incorrect predictions to lower $\rm{P}$. Strategy 2 comprehensively considered $\rm{P}$ and $\rm{R}$, achieving the best event-based and segment-based macro F1 score of $35.00\%$ and $64.32\%$ respectively. The results show it is feasible and effective to use audio tagging, a relatively simple task, to assist detection.
\begin{table}[t!]
    \centering
    \caption{The performance of audio tagging models}
    \begin{tabular}{cc}
        \toprule       
        Feature extractor &  Audio tagging macro F1[\%]\\
        \midrule  
        Backbone (trainable) &  $\bm{77.24}$ \\
        1D-DETR Backbone (fixed) & $74.12$  \\
        SEDT-AQ Backbone (fixed) & $75.44$ \\
        \bottomrule  
    \end{tabular}
    \label{tab:backbone}
\end{table}
\subsubsection{One-to-many matching strategy}
We explore the impact of $\epsilon$ and $\alpha$ on the performance of models with different fusion strategies. 
As shown in Figure \ref{parameter}, we vary the location cost threshold $\epsilon$ from $-2$ to $6$ in increments of $0.5$, for the location cost of almost all predictions is in this range. As for $\alpha$, we set it as $0.5, 1, 2$, and at the same time give the results when all predictions with location cost within $\epsilon$ are retained, denoted as ``all”. Overall, we can intuitively find that models with $\alpha=0.5$ and $\alpha=1$ perform similarly, but the latter fluctuates more noticeably with the change of $\epsilon$. Both these two curves show an upward trend before $\epsilon=0.5$, and then flatten out. When $\alpha=2$, the performance of the models drop significantly, but they are still better than the models with all qualified predictions retained. The ``$\alpha=2$'' and ``all'' curves are not sensitive to $\epsilon$.

The figure suggests that it makes sense to tune $\alpha$ since there is an apparent gap between models with a small $\alpha$ and models with ``all''. However, $\epsilon$ has little effect on the models, especially when it is greater than 0.5. $\epsilon$ cannot be set too small, otherwise, it will greatly reduce the performance of models. We argue this is because when $\epsilon$ is small, only a small proportion of the predictions will be utilized to form additional matching, limiting the influence of fine-tuning. When $\epsilon$ is large, the number of predictions that meet the positioning requirements increases. If all of them are added to the matching relationship during training, different event query slots will be optimized towards the same targets. Then during inference, these event query slots will tend to give similar detection results and miss some events, which will reduce the recall of the model. Actually, the setting of $\epsilon$ and $\alpha$ is seeking a balance between generalization ability and a better fitting of the training dataset. Regarding the setting of these two parameters, our suggestion is to apply a large $\epsilon (\textgreater 0)$ with a small $\alpha (\textless 2)$. The optimal combination can be searched under this principle.

\begin{figure*}[t!]
\centering
\includegraphics[width=7.16in]{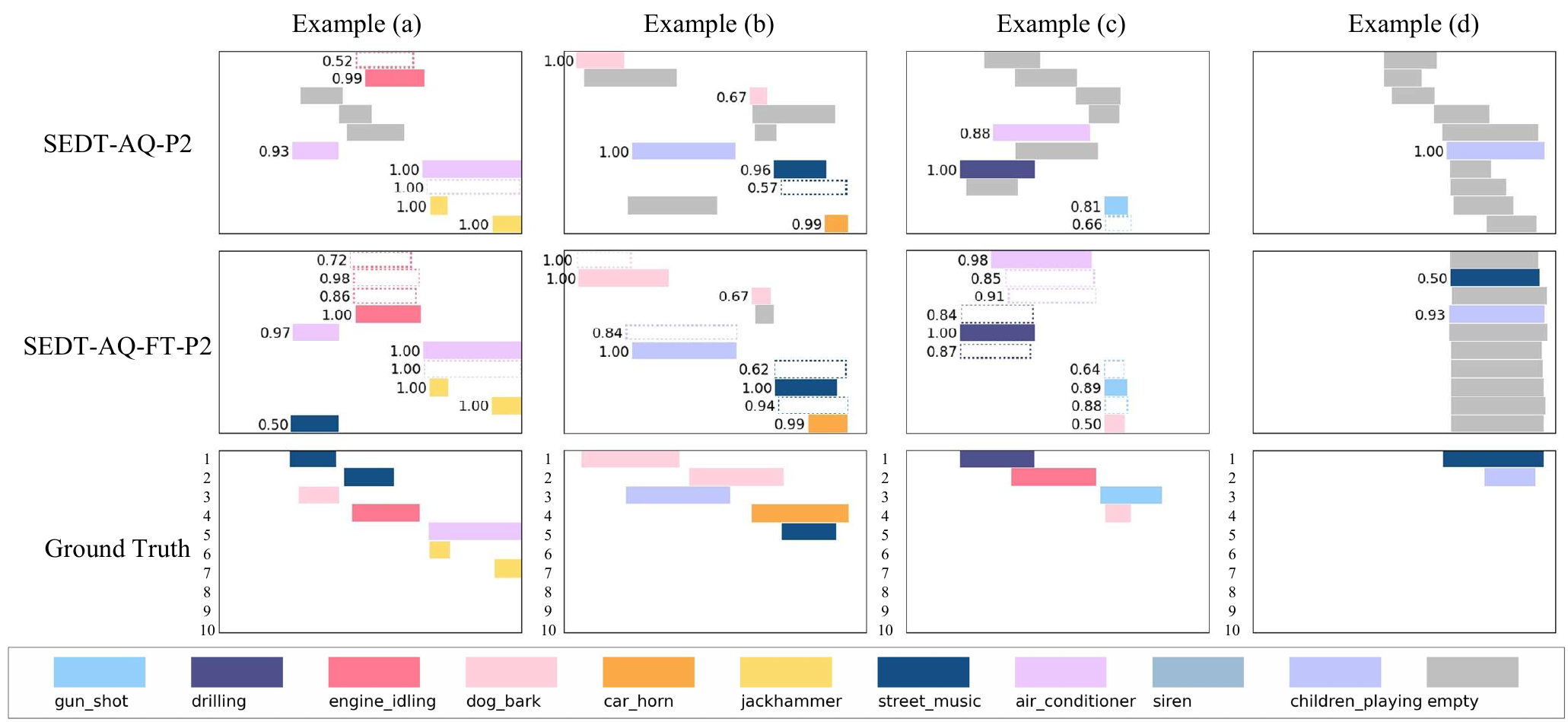} 
\caption{Visualization of predictions, where the horizontal axis represents time, and the vertical axis represents the index of predictions, numbered from 1 to 10 from top to bottom. The boxes with the dashed border represent the predictions that are deleted in the de-overlapping phase.}
\label{results}
\end{figure*}

With $\epsilon=1, \alpha=1$, We fine-tune the best model obtained in the learning stage, i.e., SEDT-AQ-P2, by the one-to-many matching method to get the final model which is referred to as SEDT-AQ-FT-P$m$, where $m$ denotes that the event-level classification fusion strategy $m$ is applied. As shown in Table \ref{tab:ablation}, SEDT-AQ-FT-P1 is $2.27\%$ higher than SEDT-AQ-P2 on the event-based macro F1 and $0.89\%$ higher on the segment-based macro F1. To show the impact of one-to-many matching on the model more intuitively, we visualize the distribution of the predictions obtained by SEDT-AQ-FT-P2 and SEDT-AQ-P2 for the same audios in Figure \ref{results}. It can be seen that after fine-tuning, the distribution of boxes is more concentrated, that is, there will be several predictions targeting the same ground truth event, leading to the reduction of results predicted as ``empty'' class. One-to-many matching also helps to predict more accurate boundaries, such as the predictions for the ground truth event ``dog bark'' with index 1 in example (b). What's more, it helps to generate correct predictions which are omitted by SEDT-AQ-P2, such as the predictions for the ground truth event ``street music'' with index 1 in example (a) and the predictions for the ground truth event ``dog bark'' with index 4 in example (c).

After fine-tuning, the rules of $\rm{P}$ and $\rm{R}$ over different strategies described above remain, but the model with the best F1 score has changed. For SEDT-AQ-P$m$, the model with strategy 2 gets the highest event-based macro F1, while after fine-tuning, strategy 1 performs best. The reason may be that the essence of fine-tuning is to improve $\rm{R}$ by mitigating predictions that are labeled as ``empty'' class, while strategy 1 focuses on improving $\rm{P}$ by deleting possible wrong predictions, their combination can complement each other to achieve the best performance. However, for Strategy 2, it already takes both $\rm{R}$ and $\rm{P}$ into account, after fine-tuning, it pays too much attention to $\rm{R}$, therefore lowering F1. 
\section{Conclusions}
In this paper, based on the characteristics of the one-dimensional signals, we present the one-dimensional Detection Transformer (1D-DETR) aiming at event-based SED, which is inspired by Detection Transformer for image object detection. We further propose the audio query branch and one-to-many matching strategy, obtaining Sound Event Detection Transformer (SEDT). To our knowledge, SEDT is the first end-to-end SED model, which gets rid of the frame-based prediction manner and generates event-level output directly. Experiments on the URBAN-SED dataset and the DCASE2019 Task4 dataset show that it can achieve competitive performance. For weakly supervised learning, SEDT relies on synthetic data to obtain localization ability. In the future, we will attempt to train SEDT to locate random cropped patches so that localization ability can be obtained without any strongly-labeled data, either real recordings or synthetic clips, realizing totally weakly supervised learning. The semi-supervised learning and unsupervised learning methods for SEDT are also our future research interests. 

\ifCLASSOPTIONcaptionsoff
  \newpage
\fi



\bibliographystyle{IEEEtran}
\bibliography{IEEEexample.bib}
%



%


\clearpage
\begin{IEEEbiographynophoto}{Zhirong Ye}
received the B.E. degree in Communication Engineering from Wuhan University, Wuhan, China, in 2019. She is currently pursuing an M.E. degree in Computer Science and Technology in Institute of Computing Technology, Chinese Academy of Sciences, Beijing, China. Her research interest includes audio signal processing and machine learning.
\end{IEEEbiographynophoto}
\begin{IEEEbiographynophoto}{Xiangdong Wang}
is an associate professor in Institute of Computing Technology, Chinese Academy of Sciences, Beijing, China. He received Doctors degree in Computer Science at Institute of Computing Technology, Chinese Academy of Sciences, Beijing, China, in 2007. His research field includes human computer interaction, speech recognition and audio processing.
\end{IEEEbiographynophoto}
\begin{IEEEbiographynophoto}{Hong Liu}
is an associate professor in Institute of Computing Technology, Chinese Academy of Sciences, Beijing, China. She received her Doctors degree in Computer Science at Institute of Computing Technology, Chinese Academy of Sciences, Beijing, China, in 2007. Her research field includes human computer interaction, multimedia technology, and video processing.
\end{IEEEbiographynophoto}
\begin{IEEEbiographynophoto}{Yueliang Qian}
is a professor in Institute of Computing Technology, Chinese Academy of Sciences, Beijing, China. He received his Bachelors degree in Computer Science at Fudan University, Shanghai, China in 1983. His research field includes human computer interaction and pervasive computing.
\end{IEEEbiographynophoto}
\begin{IEEEbiographynophoto}{Rui Tao}
 is a researcher in Toshiba China R\&D Center, Beijing, China. She received Bachelors degree in Communication Engineering at Tianjin University of Commerce, Tianjin, China, in 2017. She is currently pursuing an M.E degree at University of Chinese Academy of Sciences. Her research field includes human computer interaction and speech recognition.
\end{IEEEbiographynophoto}
\begin{IEEEbiographynophoto}{Long Yan} 
is a researcher in Toshiba China R\&D Center, Beijing, China. He received Doctors degree in signal and information processing at Beijing University of Posts and Telecommunications, Beijing, China, in 2005. His research field includes human computer interaction, speech recognition and audio processing.
\end{IEEEbiographynophoto}
\begin{IEEEbiographynophoto}{Kazushige Ouchi}
received B.E. and M.E. degrees in 1996 and 1998 from Waseda University, Tokyo, Japan and then affiliated with Toshiba Corporation. He received Ph.D. from Waseda University in 2017. He used to be a chief research scientist at Toshiba Corporate R\&D Center and is currently a vice president of Toshiba (China) Co., Ltd., and a director of Toshiba China R\&D Center. He is engaged in research on human interface using various AI technologies. He is a board member of the Academy of Human Informatics and is a senior member of the Information Society of Japan (IPSJ). He has been awarded several prizes: IPSJ Nagao Special Research Award, IPSJ Yamashita SIG Research Award, the Best Paper Award from Human Interface Society, and so on.
\end{IEEEbiographynophoto}





\end{document}